\documentclass[journal,10pt]{IEEEtran}

\usepackage{color}
\usepackage{epsfig}
\usepackage{epstopdf}
\usepackage{cite}
\usepackage[cmex10]{amsmath}
\usepackage{array}
\usepackage[footnotesize]{subfigure}
\usepackage{fixltx2e}
\usepackage[version=3]{mhchem}
\usepackage[T1]{fontenc}

\begin{document}
%
\title{Lightweight Joint Compression-Encryption-Authentication-Integrity Framework Based on Arithmetic Coding}

\author{\IEEEauthorblockN{Alaa Eldin Rohiem Shehata and Hassan Yakout El-Arsh\\}
\IEEEauthorblockA{Department of Communications, The Military Technical College\\
\quad Email: alaa\_rohiem@yahoo.co.uk, hassan.yakout@ymail.com }
}


\maketitle
\setlength{\textfloatsep}{10pt}
\vspace{-10pt}
\begin{abstract}
Arithmetic Coding is an efficient lossless compression scheme applied for many multimedia standards such as JPEG, JPEG2000, H.263, H.264 and H.265. Due to nonlinearity, high error propagation and high error sensitivity of arithmetic coders, many techniques have been developed for extending the usage of arithmetic coders for security as a lightweight joint compression and encryption solution for systems with limited resources. Through this paper, we will describe how to upgrade these techniques to achieve an additional low cost authentication and integrity capabilities with arithmetic coders. Consequently, the new proposed technique can produce a secure and lightweight framework of compression, encryption, authentication and integrity for limited resources environments such as Internet of Things (IoT) and embedded systems. Although the proposed technique can be used alongside with any arithmetic coder based system, we will focus on the implementations for JPEG and JPEG2000 standards.
\end{abstract}


%
\IEEEpeerreviewmaketitle

\indent
\section{Introduction}\label{sec:intro}

Through the past few years, combining both coding and encryption in a single algorithm to reduce the complexity is a new tempting approach for securing data during transmission and storage \cite{nonlinear1, nonlinear2, nonlinear3, Wu:2005, Zhou:2007, Nagaraj:2009, Grangetto:2006, ME}. This new approach aims to extend the functionality of coding algorithms to achieve both coding and encryption simultaneously in a single process without an additional encryption stage like many other traditional schemes such as Pretty Good Privacy (PGP) \cite{PGP} and Internet Protocol Security (IPsec) \cite{IPSEC1, IPSEC}.

According to \cite{Grangetto:2006,ME}, employing the new combined simultaneous compression-encryption approach highly reduces the required resources for encryption (computational and power resources). Also, the new approach preserves all available standard features which are not available when applying traditional encryption schemes, such as progressive transmission for JPEG2000 \cite{JPEG:2004} (also available for JPEG \cite{JPEG:1994}) and the random access feature (also called compressed domain processing) in JPEG2000. Furthermore, the new approach achieves more features and capabilities over traditional encryption schemes such as multilevel security access. The most attracting target for this new approach is the arithmetic coder.

Arithmetic coder is a lossless entropy coder used for most widespread multimedia coding standards as a last compression stage \cite{JPEG:1994, JPEG:2004, H.263, H.264 ,H.265,JBIG,AAC:2009}. This is due to its higher compression efficiency than traditional Huffman coder \cite{arith_eff}. Arithmetic coder is included in JPEG image codec \cite{JPEG:1994} and H.263 video codec \cite{H.263} as an alternative option for Huffman coder. For more recent multimedia standards, which require more compression performance like JPEG2000 \cite{JPEG:2004} and JBIG \cite{JBIG} image codecs, H.264 \cite{H.264} and H.265 \cite {H.265} video codecs, arithmetic coder is mandatory.

In this paper, a lightweight authentication and integrity capabilities is proposed exploiting the \textbf{\textit{nonlinear properties}} of the arithmetic coder. The following section describes some detailed information about various properties for arithmetic coders, which is necessary to explain the rest of the paper. Section \ref{sec:review} illustrates the previous work related to the proposed technique. Section  \ref{sec:proposed} provides a complete explanation of the proposed technique and introduces comparisons between the proposed technique and related works. Finally, conclusions and contributions of this paper are summarized in Section \ref{sec:concs}.

\section{Arithmetic coding}\label{sec:ac}
Arithmetic Coding is a variable-length entropy coding technique used for lossless data compression. First approaches to arithmetic coder were already invented in 60's \cite{Arith:63}. Then, arithmetic coding was able to gain more interest in the 80's due to its high compression efficiency compared to the well-known Huffman coding algorithm \cite{arith_eff} because the arithmetic coder overcomes the constraint that each symbol has to be coded by an integer number of bits as will be described with example in subsection \ref{sec:ac:adv}. Nowadays, there are many applications using arithmetic coding as described in section \ref{sec:intro}.

\subsection{Simplified Example}\label{sec:ac:simple}
\begin{figure}
  \centering
  \includegraphics[width=80mm]{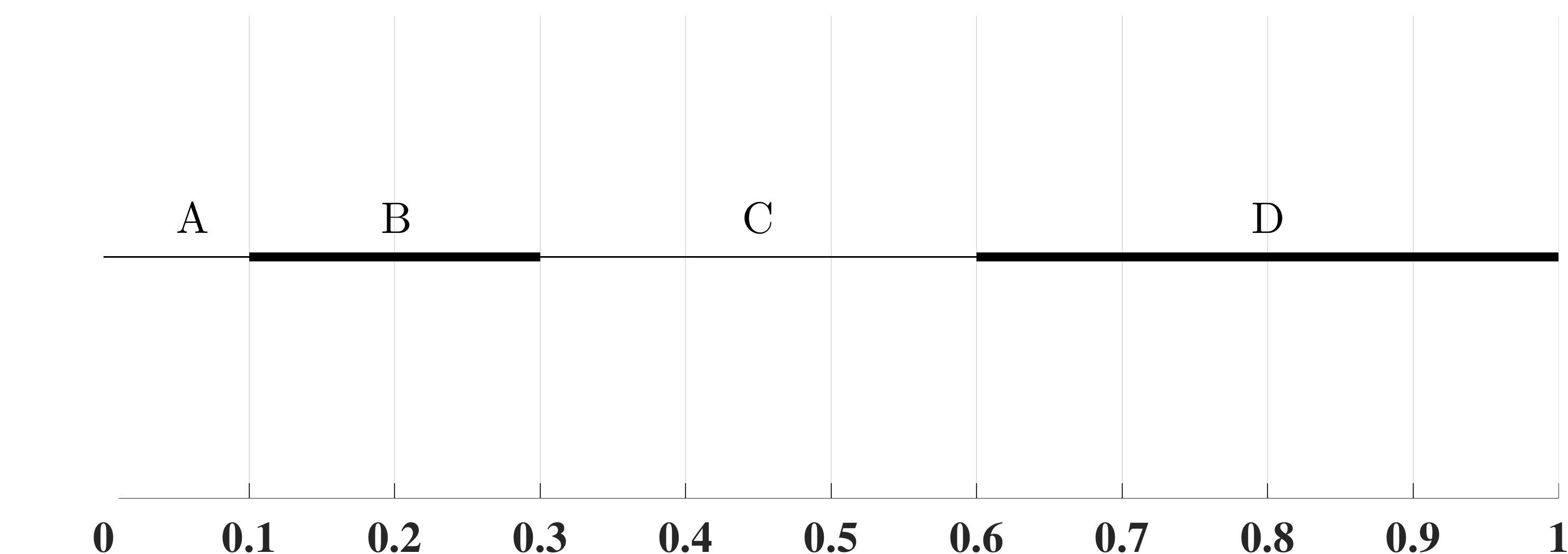}
  \caption{Probability Map.}
  \label{FIG:ArithMAP}  
\end{figure}

The following example describes the process of arithmetic encoding and decoding. Assuming discrete-memoryless source with four symbols $\{A,B,C,D\}$ with probabilities $\{P_A=0.1, P_B=0.2, P_C=0.3, P_D=0.4\}$. The encoder creates a model called {\textit{\textbf{Probability Map}}} starting from $0$ to $1$ as in Fig.\ref{FIG:ArithMAP}. 

Generally, the algorithm for arithmetic encoding works as follows \cite{Arith:63}:

\begin{enumerate}
  \item Coding starts with a \underline{\textbf{current interval}} [L,H] which is initialized to [0,1]. In this context, (L) and (H) are acronyms for (Low) and (High) respectively.
  \item For each symbol of the uncompressed data to be coded, two steps are performed:
  \begin{enumerate}
    \item The current interval is divided into subintervals, one subinterval for each symbol. The size of a symbol's subinterval is directly proportional to the symbol's predicted probability.
    \item The encoder selects the subinterval assigned to the symbol that actually occurs, and mark it as the new \underline{\textbf{current interval}}.
  \end{enumerate}
  \item The width of the final subinterval is equal to the product of the probabilities of the actually occurred symbols. Then, the encoder chooses any point from this final interval and that point will be the output of the encoding process.
\end{enumerate} 

\begin{figure*}
  \centering
  \includegraphics[width=\textwidth,height=60mm]{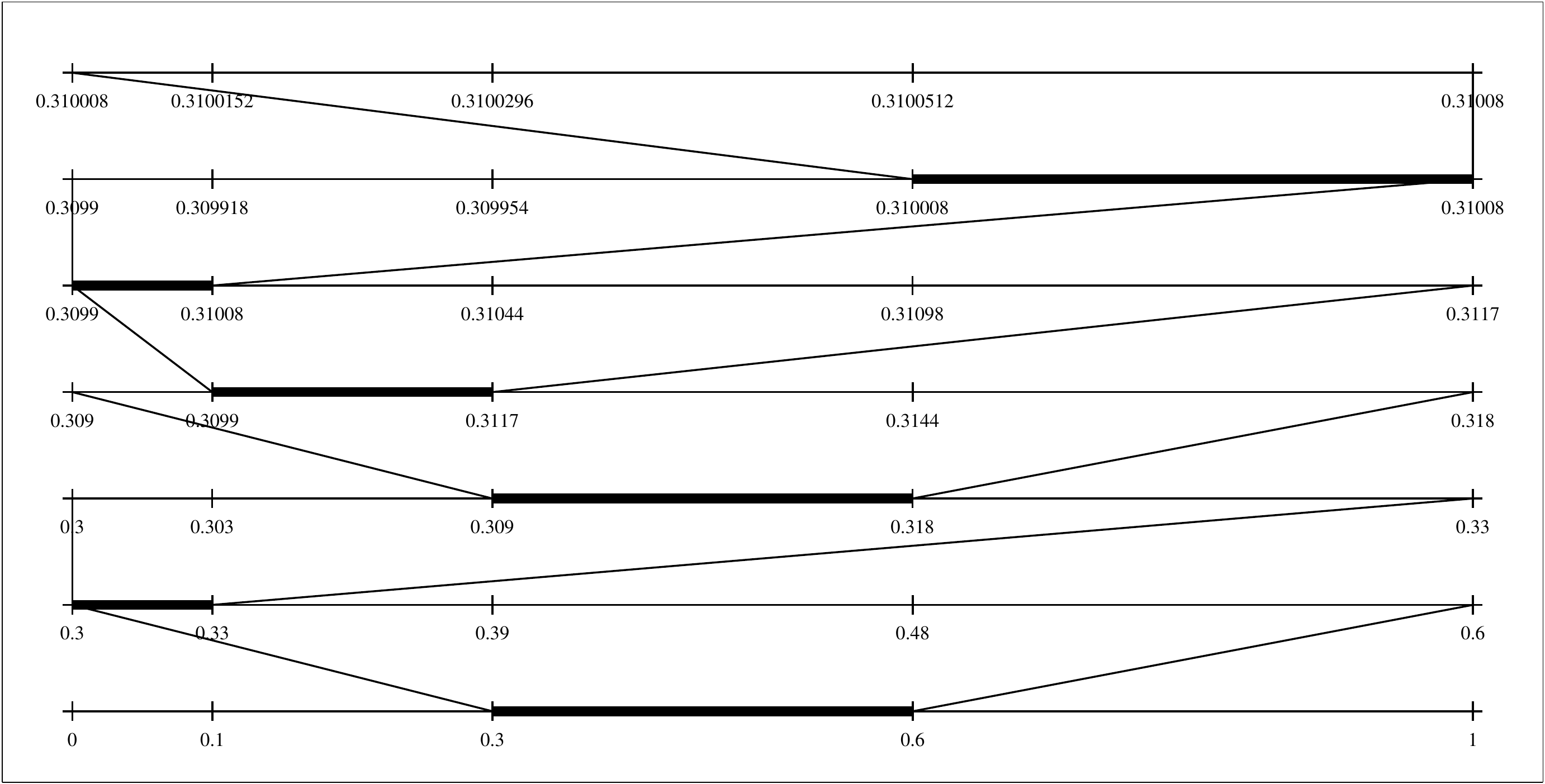}
  \caption{Function for the arithmetic coder.}
  \label{FIG:ArithPlot}  
\end{figure*}

Fig.\ref{FIG:ArithPlot} shows the above procedure for our simple example when coding a stream of data $S=\{CACBAD\}$. Here, the output may be any point within the latest sub-interval, so we can select $0.31003$.

Now, the arithmetic decoder receives the point $0.31003$, and the decoder must have the same {\textit{\textbf{Probability Map}}} for the encoder and must be informed how many symbols to decode. The decoder does the same as the encoder: it finds that the point $0.31003$ is located inside the third interval, then the first symbol will be $C$. Then dividing the third sub-interval [0.3,0.6] with the same probability ratios as in Fig.\ref{FIG:ArithPlot}. Now, the point $0.31003$ is located inside the first sub-interval, so the second symbol will be $A$. Doing the same until the end of the stream (the decoder knows that there are 6 symbols), the decoder will recover the coded stream $S$. 

\subsection{Advanced Example}\label{sec:ac:adv}
Subsection \ref{sec:ac:simple} describes a basic implementation for the arithmetic coder which has the following problems:
\begin{enumerate}
  \item As the stream of data goes longer and longer, the calculations become more and more complex as infinite precision is needed.
  \item There is no mechanism to optimally select the point representing the last sub-interval. This point theoretically requires infinite precision as the encoder may select a point which needs large number of bits to represent it. In the example in Subsection \ref{sec:ac:simple}, the encoder can select $0.310032232365766543$ instead of $0.31003$.
\end{enumerate}

For the first problem, two important properties should be noticed. First, coding process works with probabilities (\textit{i.e. ratios}). So, it is not obligatory to work within range [0,1] only, encoder and decoder can \textbf{rescale} the working region within [0,2] or [0,5] with the same probabilities (\textbf{Ratios}). Also, \textbf{shifting} the range to be [1,2] or [7,8] with the same ratios has no effect on the arithmetic coding efficiency in case the decoder at the other side uses the same conditions. Moreover, encoder and decoder can do both shifting and rescaling at the same time (for example, working in a range like [1,4] or [5,10]). The only two restrictions here is that both decoder and encoder must operate at the same manner and the probability region assigned for each symbol must match the actual symbol probability.

The second property can be concluded from Fig.\ref{FIG:ArithPlot}. It is clear that over the coding process, for each sub-interval, the upper probability value does not ever increase (may be stationary or decreasing, but never increase) and the lower probability value does not ever decrease (may be stationary or increasing but never decrease). So, upper and lower bounds are converging across encoding or decoding processes. Thus, as the lower and upper probability values start to converge, it is likely for both to have identical MSBs (most significant bits). Once upper and lower bounds have matching MSBs, the MSBs will continue to be matched till the end of the encoding or decoding processes. From that point on, MSBs will be fixed across all following coding stages. Shifting the matched MSBs out to the encoded data stream, 
and append a 0-bit for the LSBs of the lower limit and upper limit will double the current sub-interval and will reduce the complexity of calculations. For example:

\begin{enumerate}
\item Assuming Lower Range = \textcolor{red}{\underline{1}}011010101
\item Assuming Upper Range = \textcolor{red}{\underline{1}}101001000
\item Shift out the MSB and write it to the encoded output:
\begin {itemize}
\item Lower Range = 011010101\_
\item Upper Range = 101001000\_
\end{itemize}
\item Adding the new LSB:
\begin {itemize}
\item Lower Range = 0110101010
\item Upper Range = 1010010000
\end{itemize}
\end{enumerate}
So, encoding and decoding processes can be performed within a fixed binary precision, and hence the first problem is solved.

The second problem is easy to solve as the process to find the optimum binary representation for the width of the final region is simple as shown by next numerical example:
\begin{enumerate}
  \item If the final region is $\geq 0.5$, only one bit is needed to represent it (As $\{0.0 , 0.5\} \in \{0.0_2 , 0.1_2\}$) as this region can't be positioned within the interval $[0,1]$ without passing across any of $\{0.0 , 0.5\}$.
  \item If the final region is $\geq 0.25$, only two bits are needed to represent it (As $\{0.0 , \allowbreak 0.25 , \allowbreak 0.5 , \allowbreak 0.75\} \in \{0.0_2 , \allowbreak 0.01_2, \allowbreak 0.01_2 , \allowbreak 0.11_2\}$).
  \item If the final region is $\geq 0.125$, only three bits are needed to represent it (As $\{0.125 ,\allowbreak 0.25 , \allowbreak 0.375 , \allowbreak 0.5 , \allowbreak 0.625 , \allowbreak 0.75 , \allowbreak 0.875\} \in \{0.0_2 , \allowbreak 0.001_2 , \allowbreak 0.010_2 , \allowbreak 0.011_2 , \allowbreak 0.100_2 ,\\ \allowbreak 0.101_2 , \allowbreak 0.110_2 , \allowbreak 0.111_2\}$). 
\end{enumerate}
The width of the final region is the result of multiplication for the probabilities of all symbols included in the coded stream \cite{Arith:63}, as explained in Fig.\ref{FIG:ArithPlot}. Hence, the minimum number of bits ($B_{min}$) required for representing region length ($l_n$) is:
\begin{equation}\label{EQ:BMIN}
  B_{min} = \lceil -\log_2 (l_n) \rceil  \qquad \text{bits}
\end{equation}
Note that $B_{min}$ for equation (\ref{EQ:BMIN}) is exactly the total amount of information contained within the coded bit-stream \cite{couch:2012}. So, the represented bit-stream here is the optimum (lowest possible length) representation. This is why the arithmetic coder achieves $5-10\%$ better compression than Huffman coders as stated in \cite{arith_eff}.

\subsection{Arithmetic coder for JPEG and JPEG2000}
While \textit{MQ-}coder is the standard implementation of the arithmetic coder used for JPEG2000 standard format \cite{JPEG:2004}, the \textit{QM-}coder is the one for JPEG \cite{JPEG:1994}. The concepts and ideas are the same for both \textit{MQ-}coder and \textit{QM-}coder. Consequently, one of them, \textit{QM-}coder, will be described here. 

\textit{QM-}coder \cite{JPEG:1994} is a multiplication-free adaptive binary arithmetic coder that codes binary data streams with simple lookup-table-based operations. In order to avoid complex multiplications and scaling calculations, the \textit{QM} encoder specifies a discrete-probability table contains 113 states as described in Table \ref{tab:MQ}. The only significant implementation difference between \textit{MQ-}coder and \textit{QM-}coder is that the probability table of \textit{MQ-}coder contains 47 states.

\begin{table}
\centering 
\abovecaptionskip 5pt
  \caption{$QM$-table for JPEG}\label{tab:MQ}
\begin{tabular}{ | c | c | c | c | c | c |}
\hline
  Index   & $Q_e$ Value          & NMPS   & NLPS  & Switch     \\ \hline
  0       & $0 \times 5A1D$      & $1$    & $1$   & $1$        \\ \hline
  \vdots  &\vdots                &\vdots  &\vdots &\vdots       \\ \hline
  $i$     &$Q_i$                 & $M_i$  &$L_i$  &$S_i$      \\ \hline
  \vdots  &\vdots                &\vdots  &\vdots &\vdots     \\ \hline
  112     & $0 \times 59EB$      & 112    & 111   &    1      \\ \hline
\end{tabular}
\end{table}

Each row for the $QM$-coder's table is considered a state representing a different probability map which, due to the binary nature of $QM$-coder, can be modeled simply by the probability of the Less Probable Symbol (LPS) denoted by $Q_e$. According to Table \ref{tab:MQ}, the probability of the (MPS) and (LPS) will be $1-Q_e$ and $Q_e$ respectively.

The \textit{QM-}coder starts at an initial table entry (state) $i$, which may be modified through the coding process according to the entries NMPS (Next index after MPS ) and NLPS (Next index after LPS). During the coding process, the encoder decides whether a received input bit is coded as MPS or LPS. Thus, the next state will be either $M_i$, or $L_i$ respectively. However, if the Switch flag bit equals to $1$, the coder exchanges the value for MPS and LPS.

\subsection{Avalanche effect for arithmetic coder}\label{sec:arith:avalanche}
The avalanche effect for the arithmetic coder is an important criteria for using the arithmetic coder for security. According to \cite{nonlinear1, nonlinear2, nonlinear3}, arithmetic coder is characterized by high error sensitivity and error propagation properties. Furthermore, it is proven by \cite{Nagaraj:2009} that any arithmetic coder can be considered a chaotic random generator with proven cryptographic nonlinear properties. Moreover, a practical experiment is described in \cite{stat} uses the NIST's statistical test tool \cite{NISTSTS} to support these cryptographic properties. Consequently, this means that any change in the input bit-stream for the encoder/decoder side (even in a single bit), leads to a huge avalanche effect for the all the following encoded/decoded output bit-stream. The following example demonstrates these properties.

\begin{figure*}
  \centering
  \includegraphics[width=\textwidth,height=65mm]{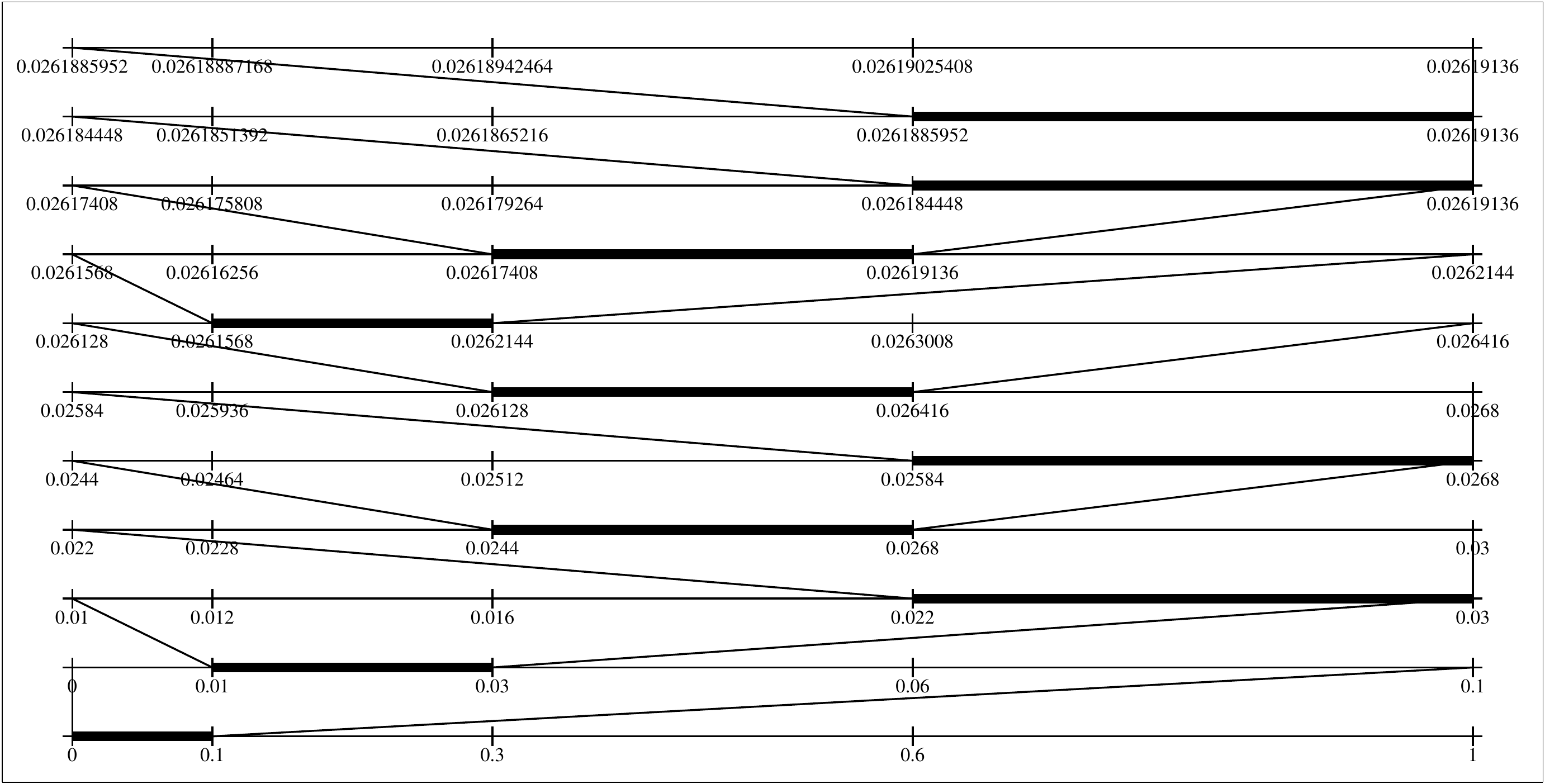}
  \caption{Normal Arithmetic coding process.}
  \label{FIG:Arith_err_normal}
\end{figure*}

\begin{figure*}
  \centering
  \includegraphics[width=\textwidth,height=65mm]{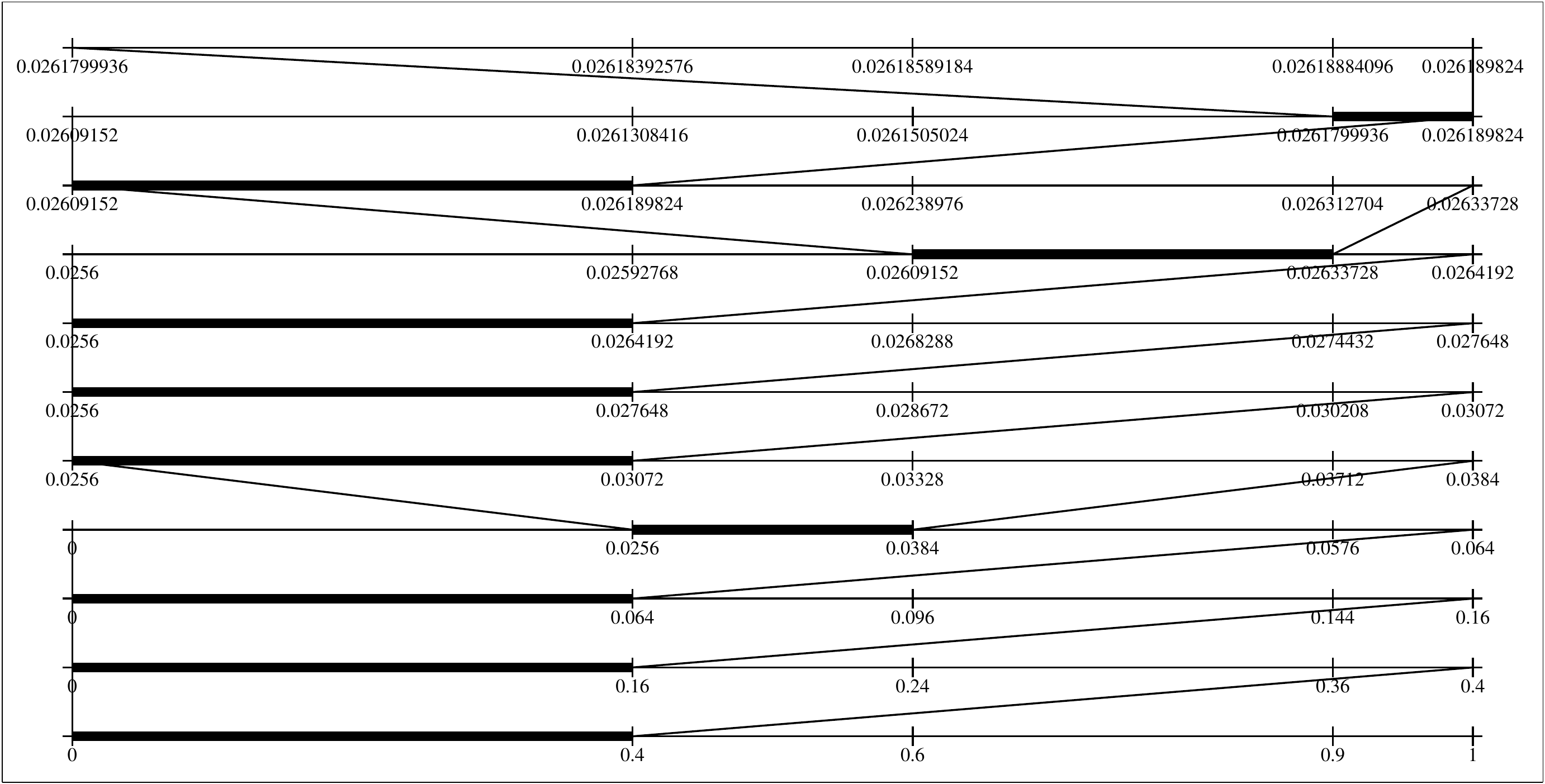}
  \caption{First type of errors.}
  \label{FIG:Arith_err_1}
\end{figure*}

Using the same discrete-memoryless source in subsection \ref{sec:ac:simple}, when coding the message $\{ABDCDCBCDD\}$, the detailed coding process and final coding result will be as in Fig.\ref{FIG:Arith_err_normal}. According to Fig.\ref{FIG:Arith_err_normal}, the point $0.026189424$ can be used as a result for coding the message. If the decoder has the same probability map, the decoder functional diagram will be the same as in Fig.\ref{FIG:Arith_err_normal} and the decoded message will be identical to the original message. Now, If the decoder uses a wrong probability map (for example $\{0.4,0.2,0.3,0.1\}$ for symbols $\{A,B,C,D\}$ respectively), the decoder function diagram will be as in Fig.\ref{FIG:Arith_err_1} and the decoded message will be $\{AAABAAACAD\}$, which is different from the original message by $70\%$. Let us name this type of errors as \textbf{\textit{First type of errors}} in which the decoder has a different probability map. This type of errors also appears even when changing the order of the symbols without changing the probability values of any symbol.

\begin{figure*}
  \centering
  \includegraphics[width=\textwidth,height=65mm]{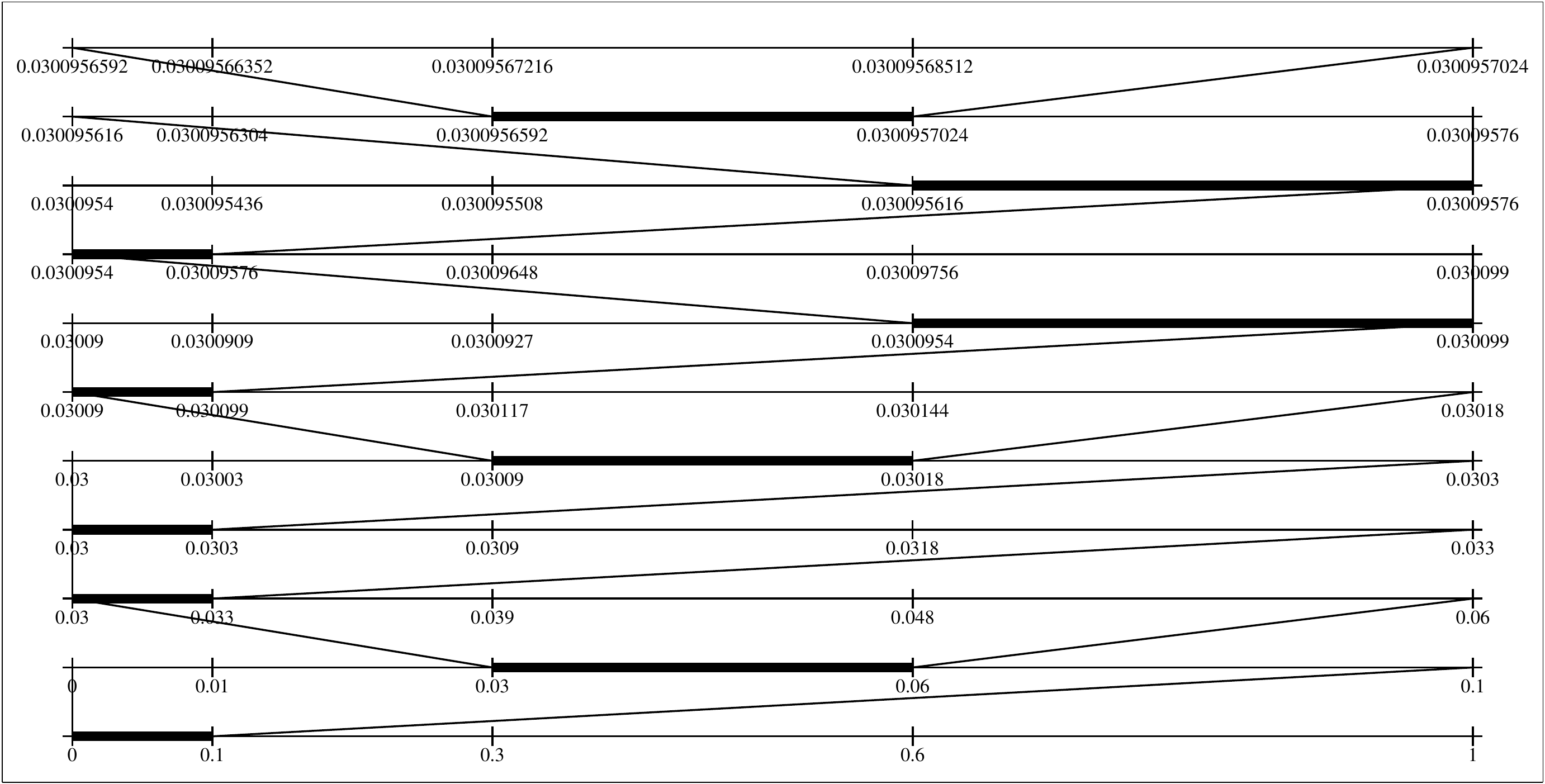}
  \caption{Second type of errors.}
  \label{FIG:Arith_err_2}
\end{figure*}

Another type of errors, called \textbf{\textit{Second type of errors}}, in which the received point has been changed. The coded message is $0.026189424$ which corresponding to the binary $0000011\textbf{\underline{0}}1011010001011001101000100101$. Now, changing it to be $0000011\textbf{\underline{1}}1011010001011001101000100101$ with a single bit error which leads to start at point will be $0.030095674635959$. Here, the decoder will have a functional diagram as in Fig.\ref{FIG:Arith_err_2} and the recovered message will be $\{ACAACADADC\}$ with $80\%$ errors. 

This type of errors when applied to any arithmetic coder, it affects only all the symbols following the bit error. If the error is in the first bit, the whole message will be decoded incorrectly, but if the error affects only the last \underline{\textit{usable}} bit (i.e. not the padding bits), the whole message will be coded correctly except the last symbol.

\subsection{Adapting arithmetic coder for security}\label{sec:ac:arith_sec}
Two types of errors for the arithmetic coder had been described in the previous subsection. The first type of errors can be used \textit{\textbf{Inside}} the arithmetic coding stage without any additional encryption stage (i.e. without other additional processing). This makes the arithmetic coder performs both compression and encryption simultaneously \cite{Grangetto:2006,ME}. Besides, this type of errors can be used also for error detection \cite{nonlinear2} and correction \cite{JPWL}, which will be described in section \ref{sec:review}.

The second type of errors can be used in tandem with traditional digital signature algorithms to sign a small part of the bit-stream to achieve low complexity integrity and authentication capabilities, which is the main idea of the proposed work, described in section \ref{sec:proposed}. This type of errors is used \textit{\textbf{Outside}} the arithmetic coding stage. 

\begin{figure*}
  \centering
  \includegraphics[width=\textwidth,height=65mm]{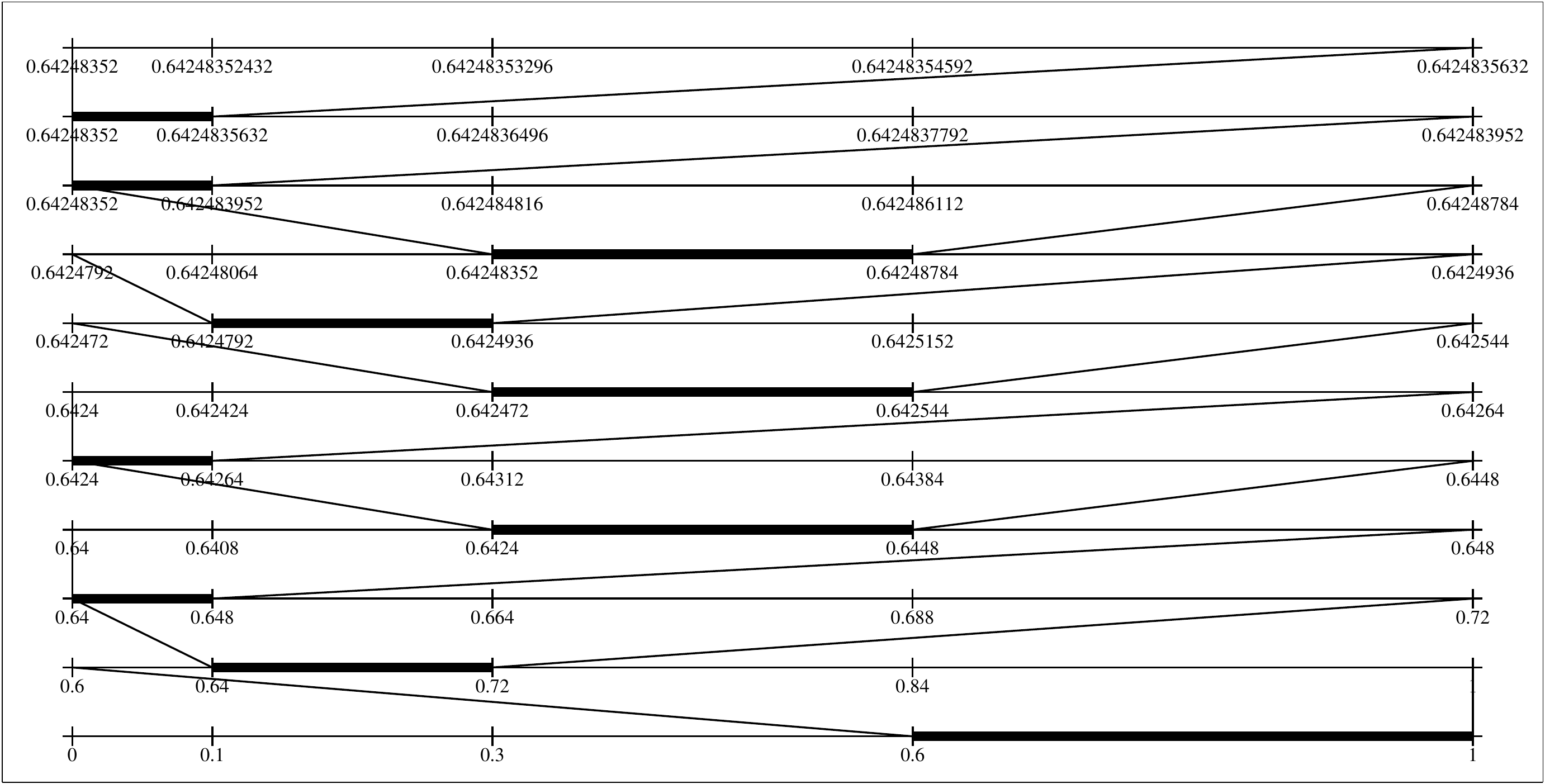}
  \caption{Disadvantage of the first type of errors.}
  \label{FIG:Arith_err_expansion}
\end{figure*}

When using the first type of errors, the following criteria must be carefully taken into account. Referring to Fig.\ref{FIG:Arith_err_normal}, the width of the final interval will be ($0.27648\times 10^{-5}$), but when applying the first type of errors at the encoder (to achieve encryption) by making the probability model to be $\{0.1,0.2,0.3,0.4\}$ for the symbols $\{D,B,C,A\}$ respectively (i.e. swapping both symbols $A$ and $D$), as described in Fig.\ref{FIG:Arith_err_expansion}, the width of the final interval will be ($0.00432\times 10^{-5}$), which equals ($\frac{1}{64}$) of the original interval width without encryption. Thus, according to equation (\ref{EQ:BMIN}), the output bit-stream will be expanded by an extra 6-bits as this type of errors reduces the compression efficiency because it doesn't maintain the recommended statistical model. Besides, it should be noted that as the message goes longer, the number of needed extra bits will be increased. This is an important criteria to be considered as it may cause \textit{\underline{expansion}} instead of compression as described in \cite{ME} with a practical example.

To achieve security without sacrificing the compression efficiency, the probability model should be maintained without any modifications. This can be done by applying the same permutations for both the probability model and the symbols' order (i.e. Maintain the same probability region width for each symbol). This doesn't affect the compression efficiency as the compression efficiency doesn't be affected by how we order the symbols within the probability model. This criteria has been utilized for joint compression and encryption in \cite{Grangetto:2006,ME}.

\begin{figure*}
  \centering
  \includegraphics[width=\textwidth,height=65mm]{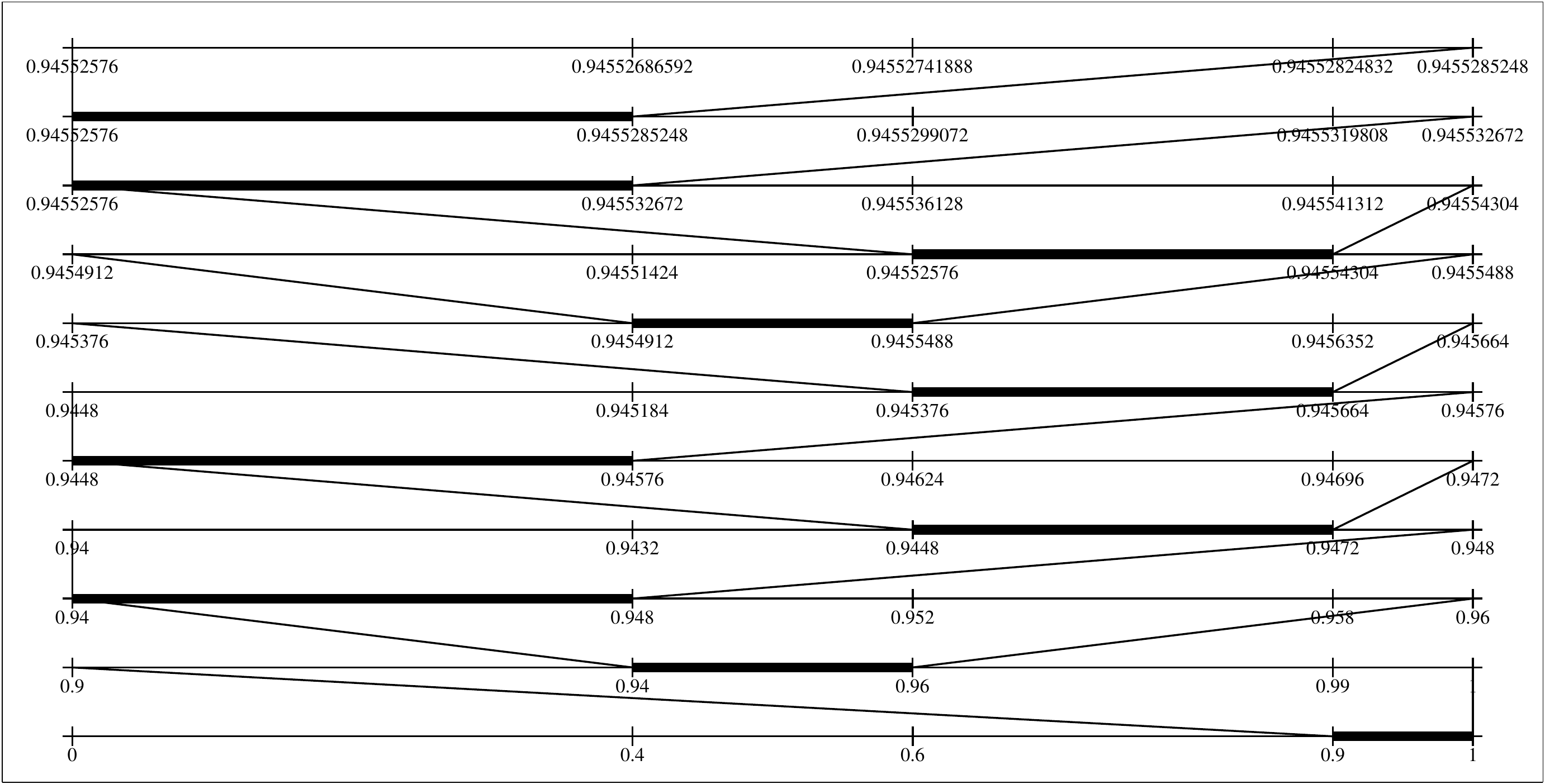}
  \caption{First type of errors without losing the compression efficiency.}
  \label{FIG:Arith_err_3}
\end{figure*}

Now, applying this type of errors for Fig.\ref{FIG:Arith_err_normal} by making the probability model to be $\{0.4,0.2,0.3,0.1\}$ for the symbols $\{D,B,C,A\}$ respectively, the obtained results described in Fig.\ref{FIG:Arith_err_3}. The width of the final interval for both Fig.\ref{FIG:Arith_err_normal} and Fig.\ref{FIG:Arith_err_3} will be ($0.27648\times 10^{-5}$). So, the compression efficiency is maintained and additional security benefit is gained.

The permutation of the symbols can be done with a key to achieve \textit{\textbf{Encryption}} \cite{Grangetto:2006,ME}. Assuming $N$ is the number of all possible symbols, the key space here will be equal to $N!$ and the more possible symbols the more the key space.

\section{Review of previous work}\label{sec:review}
\subsection{Forbidden Symbol}\label{subsec:fs}
Forbidden Symbol(s) \cite{forbidden:1997,forbidden:2000,Grangetto:2003} is a technique utilizing the nonlinearity, high error propagation and high error sensitivity of arithmetic coder for extending the arithmetic coder usage for error detection, as described in \cite{nonlinear2}, in tandem with compression and encryption. Furthermore, according to \cite{JPWL}, error correction capabilities are also achievable. This is done by inserting one or more \underline{\textit{dummy}} symbols within the probability map. This technique can be considered an application for the first type of errors, described in subsections \ref{sec:arith:avalanche} and \ref{sec:ac:arith_sec}.

Dummy symbol(s) assigned a relatively small probability value(s) (\textit{i.e.} small region(s) in the probability map). The cost for this additional feature is reducing the compression efficiency as discussed below.

According to subsection \ref{sec:ac:arith_sec}, assigning an incompatible probability values within the probability map reduces the compression efficiency. As discussed in \cite{calculations,Grangetto:2003}, assuming the total probability of the forbidden symbol(s) is ($\varepsilon$), then the actual used probability map will be ($1 - \varepsilon $) for each coding iteration. Hence, the following equation describes the amount of redundant bits $R$ per each coding step:
\begin{equation}\label{EQ:R}
R = -\log_2 (1-\varepsilon )
\end{equation}
In addition to equation (\ref{EQ:R}), assuming the length of the uncompressed stream is $N$ symbols, then the following equation calculates the total additional length when coding this uncompressed stream with an arithmetic coder applying the concept of forbidden symbol:
\begin{equation}\label{EQ:t}
L = N\times R
\end{equation}

\begin{figure}
  \centering
  \includegraphics[width=80mm,height=50mm]{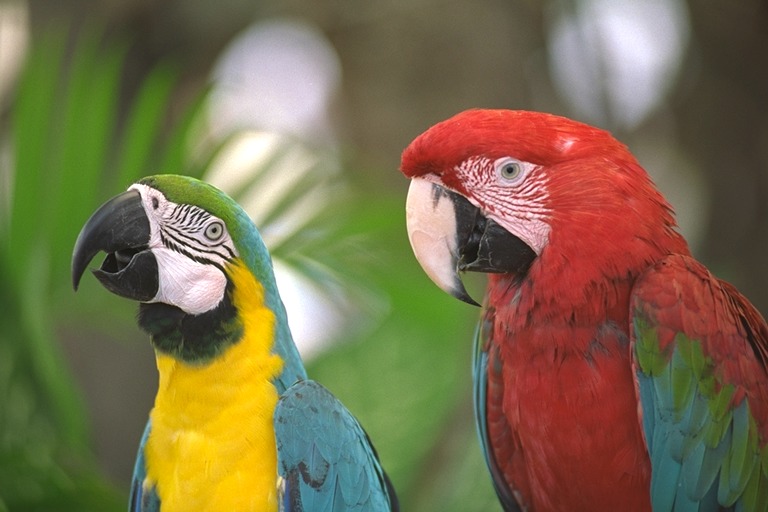}
  \caption{Sample JPEG2000 image.}
  \label{baby}  
\end{figure}

Practically, the value of $L$ can be quite small and could be ignored. To prove this, consider the \textit{MQ-}coder as an example. The minimum applicable width of the region for any symbol within the probability map of \textit{MQ-}coder is $0.000023$ \cite{JPEG:2004}, which can be the assigned width for the forbidden symbol ($\varepsilon$). Hence, as the typical code-block length is $64\times 64$ bytes \cite{Acharya:2005,JPEG:2004}, so the maximum length of the bit-stream to be coded is $2^{15}$ bits ($64\times 64\times 8$). Assuming the least possible size of the total probability map which equals to 0.75 instead of 1 \cite{JPEG:2004} (\textit{i.e.} $\varepsilon=0.000023\div 0.75$), then applying equation (\ref{EQ:R}) and equation (\ref{EQ:t}), an additional $1.4498$ bits are added per each code-block. Regarding the sample JPEG2000 image in Fig.\ref{baby}\cite{image} which contains 309 code blocks with total compressed size of $417068$ bytes when coding with \textit{libopenjpeg} \cite{libopenjpeg} with the default coding configurations, the maximum total additional size for such an image is $56$ bytes (\textit{i.e.} maximum compression loss is $0.0134\%$) per a single forbidden symbol.

\subsection{JPsec}\label{subsec:jpsec}
Forbidden symbol technique is an error detection and correction scheme which cannot be used as an alternative to integrity \cite{Stallings:2013}. The main difference between integrity and error detection is that the error detection can detect any modification of the data (within certain \textbf{\textit{limit}}) which is \textbf{\textit{unintended}} such as channel errors, on the other hand integrity can detect \textbf{\textit{unlimited}} number of errors including \textbf{\textit{intended}} and \textbf{\textit{unintended}}. Integrity cannot correct errors, just detect it.

Thus, another standard for securing the transmission and storage of images called JPsec (\textbf{Secure JPEG 2000}) had been designed \cite{JPEGSEC}. JPsec provides encryption, source authentication and data integrity in a \textbf{\textit{compliant}} manner. This means that all available features such as progressive transmission and compressed domain processing are not affected. Techniques described in \cite{Grangetto:2006, ME} achieves only encryption with higher efficiency than JPsec, but cannot achieve data integrity and source authentication. The proposed technique extends the scheme in \cite{ME}, which is more efficient than \cite{Grangetto:2006}, to achieve integrity and authentication besides encryption.

\section{The proposed technique}\label{sec:proposed}
\subsection{Design idea}
In the context of this paper, the term \textbf{complete data stream} (CDS) refers to a stream of uncompressed symbols (just before the arithmetic coder) to be coded using a \textit{complete cycle} of coding by the arithmetic coder. A complete cycle of coding for the arithmetic coder starts by initializing the arithmetic coder's registers and ends up by flushing the arithmetic coder registers, as described in \cite{JPEG:1994,JPEG:2004}.

For JPEG2000, A typical code-block equals to 4096 bytes \cite{JPEG:2004,Acharya:2005}, so the maximum length of CDS will be 4096 bytes. The minimum length of CDS for JPEG2000 cannot be stated here as the CDS per each code-block depends on the statistics for each code-block itself. To be more accurate, applying the practical implementation described in \cite{libopenjpeg} for the image in Fig.\ref{baby}, Table \ref{TAB:baby} gives the actual calculations for the length of CDS.

For JPEG, CDS can be much longer (as images can be coded within a single CDS). Thus, when applying the implementation of jpeg in \cite{libjpeg} for all 29 reference images in \cite{image}, Table \ref{TAB:baby1} gives a real calculations for the length of CDS.

\begin{figure}
  \centering
  \includegraphics[width=80mm,height=50mm]{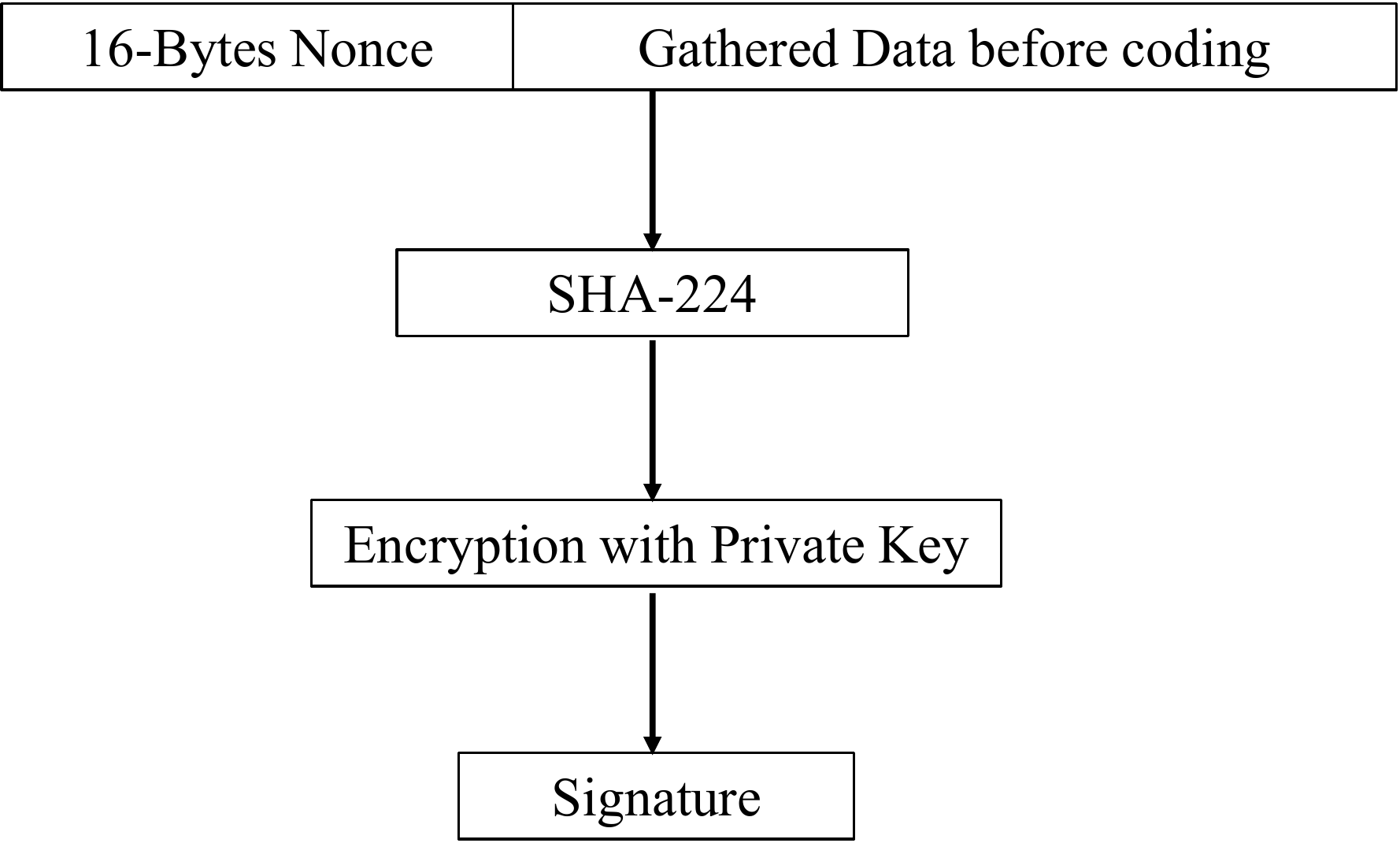}
  \caption{The proposed technique: signature at the encoder side.}
  \label{proposed:enc}  
\end{figure}

As discussed and proven in details with examples in previous sections, the arithmetic coder is characterized by high nonlinearity, high error propagation and high error sensitivity properties. Thus, by gathering a small part at the \underline{\textit{end of each CDS}} and applying any digital signature scheme to sign this gathered bytes only, integrity and source authentication can be achieved with a small cost as described in Fig.\ref{proposed:enc}. Also, the encoder must generate a unique random value (called \textit{nonce}) for each image. Nonce is appended to the gathered bytes before applying the digital signature scheme. Nonce and the final signature can be sent within the image inside the comment field for JPEG and JPEG2000 formats.

\begin{figure}
  \centering
  \includegraphics[width=80mm,height=50mm]{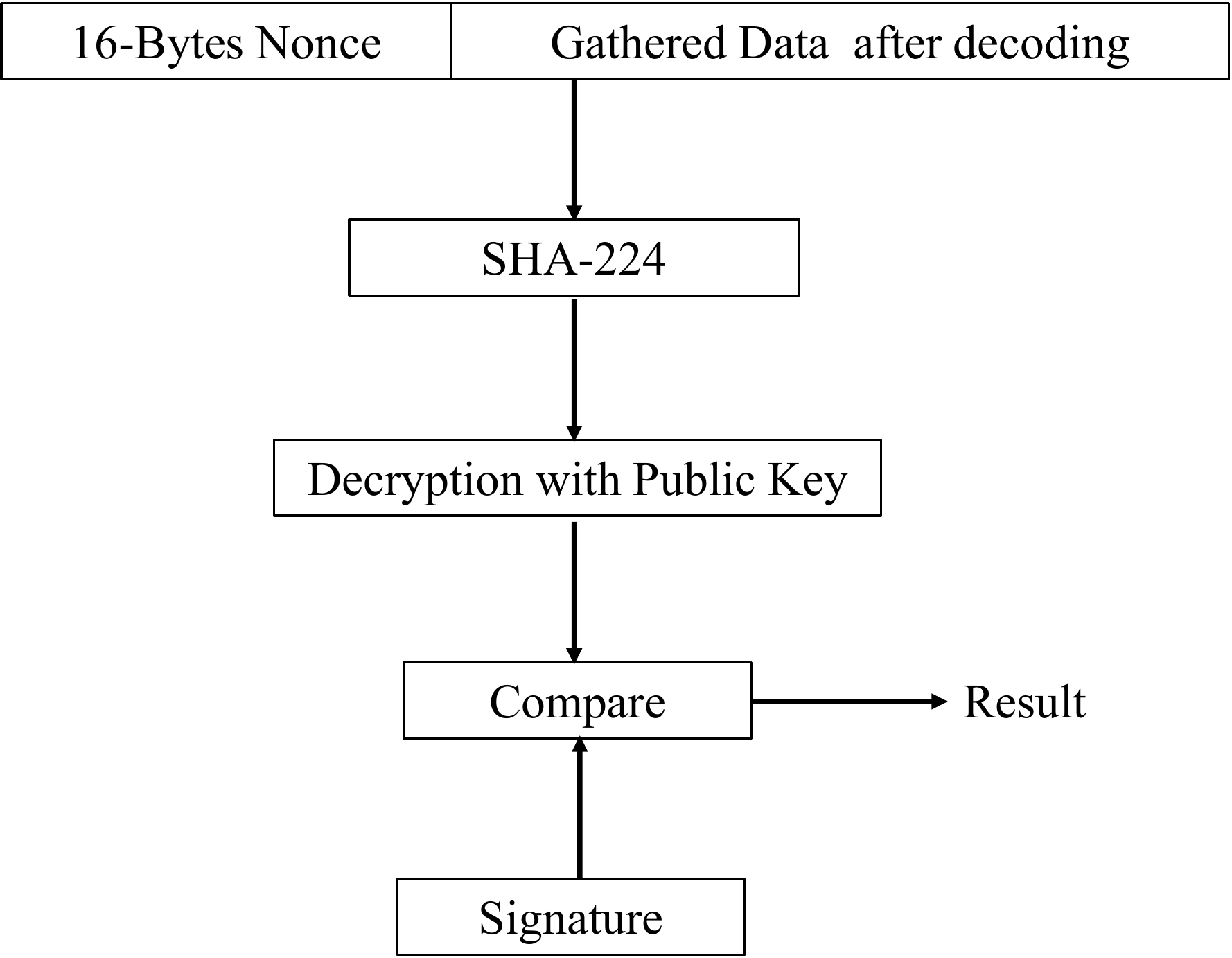}
  \caption{The proposed technique: verification at the decoder side.}
  \label{proposed:dec}  
\end{figure}

Once the signed image had been received at the decoder side, the image is first decoded by the arithmetic coder. Then, the decoder gathers data (at the end of each CDS like encoder) the and extracts Nonce, then applies the hash function to gathered data and nonce exactly the same as the encoder. After that, the decoder extracts the signature, decrypts it with the proper public key and compares this decrypted output with the signature output to verify the integrity and authenticates the source of the image as described in Fig.\ref{proposed:dec}.


\begin{table}
\centering 
  \caption{Practically calculated length (in bits) of CDS for the image in Fig.\ref{baby} using the implementation in \cite{libopenjpeg}.}
  \label{TAB:baby}
\begin{tabular}{ | c | c | c | c | }
\hline
  Max                 & Min                 & Mean                & Std. Dev.              \\ \hline
  $24298$             & $2285$              & $13759.5$              & $4468.3$                 \\ \hline
\end{tabular}
\end{table}

\begin{table}
\centering 
  \caption{Practically calculated length (in bits) of CDS for the images in \cite{image} using the implementation \cite{libjpeg} with quality factor 95.}
  \label{TAB:baby1}
\begin{tabular}{ | c | c | c | c | }
\hline
  Max                 & Min                 & Mean                & Std. Dev.              \\ \hline
  $2891308$           & $1297338$           & $1.9227\times 10^6$ & $0.4657\times 10^6$    \\ \hline
\end{tabular}
\end{table}

In this paper, SHA-224 \cite{NIST180} hash function is followed by encryption with private key algorithm (256-bits elliptic curve) \cite{NIST186} is applied to the gathered bytes concatenated with nonce.

Any other secure hash function or private key algorithm, or even another signature algorithm can be used. Using a private key algorithm assures the source authentication of the image.  

If any error occurred over the compressed stream, even in a single bit at any position, this error will diffuse through the remaining bits of the stream and will extremely affects the last few bits at the end of the recovered stream at the decoder side, and subsequently this effect can be detected by the signature algorithm. Consequently, instead of applying the signature algorithm for all bits of the coded image, only applied for a small portion of the image with the same results, thanks to the special characteristics of the arithmetic coder. This can be considered as utilizing the second type of errors described in subsection \ref{sec:arith:avalanche}.

\subsection{Implementation issues}
In the previous subsection, the minimum length of the gathered part per each CDS is 16 bytes. According to \cite{sweet32}, any block cipher or hash function uses a block length of 64-bits, is vulnerable to a \underline{\textbf{practical}} attack called \textit{Birthday paradox attack}. Thus, a minimum length for the gathered part per each CDS is 16-bytes.

Adding nonce to the gathered bits is mandatory because the last bytes of the CDS, with a high probability, are equal to zeros due to quantization \cite{JPEG:1994,JPEG:2004,Acharya:2005}. Thus, even by using a strong signature algorithm to sign the \textit{gathered data} which may be the same for more than one image is not secure. So, by adding a unique nonce as a first block of the gathered data, with minimum length of 16 bytes for each image can be considered secure.  

\subsection{Comparison with forbidden symbol technique}
As described in the previous subsections, the proposed technique utilizes the special features of the arithmetic coder, but applied \textit{\textbf{outside}} the arithmetic coder. So, it does not affect the arithmetic coder operation or the compression efficiency, unlike forbidden symbol technique. Additionally, the proposed technique has no limits for the number of detected errors as it is an \textbf{integrity} scheme, not an \textbf{error detection} scheme. Moreover, for the forbidden symbol technique, if an error occurs and the decoder did not pass through any forbidden symbol, the error will not be detected, unlike the proposed technique which utilizes a proven secure cryptographic hash function.
\subsection{Comparison with JPsec}
JPsec employs a cryptographic signature algorithm like the proposed technique, but JPsec applies this algorithm for all bytes of the CDS, unlike the proposed technique which is more efficient. Considering the JPEG2000 case, with a maximum CDS of 4096 bytes, the proposed technique is applied only for 16 bytes per each stream, which is 256 times faster than JPsec. To be more practical, considering the mean value ($12325$ bits) from Table \ref{TAB:baby}, the proposed technique is approximately 96 times faster than JPsec.

Although JPsec achieves also encryption besides integrity and source authentication, an efficient encryption technique described in \cite{ME} can be combined with the proposed technique to achieve all security services as JPsec with notably small amount of resources.

\section{Conclusions and future work}\label{sec:concs}
In this paper, a new lightweight technique is proposed to attain integrity and source authentication utilizing the arithmetic coder in a low cost manner compared to JPsec standard. Unlike the forbidden symbol technique, the proposed technique does not affect the compression efficiency of the arithmetic coder and produces a robust error detection scheme (integrity). When combining the proposed technique with \cite{ME}, an efficient and low complexity joint compression, encryption, integrity and source authentication can be achieved for systems with limited resources like IoT and embedded systems.

\bibliographystyle{IEEEtran}
\bibliography{IEEEabrv,Yakout}
\end{document}